\documentclass[a4paper,11pt]{article}
\usepackage{jheppub} 
\usepackage{physics}
\usepackage{subfigure}
\usepackage{subfigure}
\usepackage{ascmac}
\usepackage{amsthm}
\usepackage{amsmath}
\usepackage{tikz}
\usetikzlibrary{arrows.meta}
\usetikzlibrary{patterns}
\usetikzlibrary{decorations.pathmorphing}
\usetikzlibrary{decorations.text}
\usetikzlibrary{snakes}
\tikzset{>={Latex[width=2mm,length=2mm]}}

\preprint{KUNS-2932}

\title{
    String is a double slit
}

\author{Koji Hashimoto$^{*}$, Yoshinori Matsuo$^{\dagger}$, Takuya Yoda$^{*}$}
\affiliation{${}^*$Department of Physics, Kyoto University, Sakyo-ku, Kyoto 606-8502, Japan}
\affiliation{${}^\dagger$Department of Physics, Kindai University, Higashi-Osaka, Osaka 577-8502, Japan}

\emailAdd{koji@scphys.kyoto-u.ac.jp}
\emailAdd{ymatsuo@phys.kindai.ac.jp}
\emailAdd{t.yoda@gauge.scphys.kyoto-u.ac.jp}

\abstract{
    We perform
    imaging of a fundamental string
    from string scattering amplitudes,
    and show that its image is a double slit.
}

\begin{document} 
\maketitle
\flushbottom

\section{Introduction and summary}

Have you ever seen a string?
Since the discovery of string theory
as a candidate for the unifying theory,
the rich dynamics of strings
has been revealed.
For example,
strings can be reconnected and be torn apart.
They can be attached to branes
or absorbed into.
Furthermore,
strings can undergo the transition
to black holes.
The black hole-string correspondence \cite{Horowitz:1996nw,Horowitz:1997jc}
(see also \cite{Halyo:1996xe,Susskind:2021nqs})
tells us that
a single long string
tends to shrink,
if its self-gravity is turned on,
and finally it becomes a black hole
when its typical size reaches the Schwarzschild radius.
Although
it is expected that
strings
answer the microscopic degrees of freedom of black holes,
secrets are still hidden inside horizons.

If possible,
it would be worthwhile
to shoot photos of a string.
In physics and engineering,
imaging technologies are widely used
to take pictures of various objects.
Imaging technologies
allow us
to investigate the inner structures of objects
when they are hidden from the outside.
Even if their structures are
highly complicated,
they provide us
clear and intuitive ways to understand the structures.
For example,
the structure of the human body
is highly complicated
and cannot be seen directly from the outside.
However,
various imaging technologies,
such as the X-ray imaging
and the ultrasound wave imaging,
enable us to study
the functions of human organs,
and find the causes of disease.

The string is a suitable target
for applying such imaging technologies.
The inner structure of a long string
is unclear
and expected to be highly complicated.
And in fact, the traditional perturbative string theory is defined only through string scattering amplitudes, and thus if one wants to see a fundamental string itself, one needs the imaging from the scattering amplitudes, solving the inverse problem 
--- it is a reconstruction of the internal structure by the scattering wave data. 

In this paper,
as the first step,
we study tree-level scattering amplitudes
of open bosonic fundamental strings,
and reconstruct their images
from
the amplitudes.
The imaging is just a Fourier transformation, as the standard optical theory tells.
The reconstruction method
works as an eye
to look inside the scattering processes.
The results of the imaging show that
the images of the fundamental string are double-slits.

String scattering amplitudes
have a lot of zeros,
and we find that 
their pattern of brightness/darkness
observed at the spatial infinity
matches that of the standard double-slit experiment.
This is the reason why the imaging via the string scattering amplitudes results in the images of the double slits.

As a simplest example, we consider the scattering of tachyons. 
We apply the imaging method to the $s$-channel pole of the Veneziano amplitude, and find that the image is a double slit. In the large level limit of the pole mass, the coincidence with the double slit is exact.
We also apply the imaging method to the $s$-channel pole of the string decay amplitude of a highly excited string, using the amplitude obtained in \cite{Gross:2021gsj,Rosenhaus:2021xhm}. We find that the image is a set of multiple slits aligned on a straight line.

Slits in the images of the string scattering amplitudes are separated by the typical length of the fundamental string at the excited $s$-channel pole, which leads us to interpret that the slits are end points of the string. 

Since the imaging technologies are natural to be applied to string theory, more applications in various situations will reveal hidden structure of a fundamental string.

This paper is organized as follows.
In Sec.~\ref{sec:double-slit},
we review the standard double-slit experiment,
and present an imaging method
to reconstruct a double-slit image from the amplitude on a screen.
In Sec.~\ref{sec:veneziano},
we study the Veneziano amplitude,
and show that the image of a fundamental string read from the amplitude is indeed a double-slit.
In Sec.~\ref{sec:HES},
we apply the same method
to other scattering process: a highly excited string decaying into two tachyons,
and show that its image is a multi-slit.
In Sec.~\ref{sec:interpretation},
we provide possible interpretation of our results
and discuss some implications to string physics.

\section{Double-slit experiment}
\label{sec:double-slit}

In this section,
we review
the standard double-slit experiment,
and a method to reconstruct the image
from the amplitude on a screen, to demonstrate that the amplitudes actually encodes the image of the double slit.

\begin{figure}[t]
	\centering
	\begin{tikzpicture}
		\draw[->] (-2.2,0) -- (2.2,0) node[anchor=north west] {$x$};
		\draw[->] (0,-2.2) -- (0,2.2) node[anchor=south east] {$z$};
		
		\draw (0,0) circle (1.8);
		\filldraw[black] (0,0.5) circle (2pt) node[anchor=east]{$\text{S}_1$};
		\filldraw[black] (0,-0.5) circle (2pt) node[anchor=east]{$\text{S}_2$};
		\filldraw[black] ({1.8*cos(deg(pi/6))},{1.8*sin(deg(pi/6))}) circle (2pt) node[anchor=west]
		{$\text{P}$};
		
		\draw (0,0.5) -- ({1.8*cos(deg(pi/6))},{1.8*sin(deg(pi/6))});
		\draw (0,-0.5) -- ({1.8*cos(deg(pi/6))},{1.8*sin(deg(pi/6))});
	\end{tikzpicture}
	\caption{Double-slit is put at S${}_1$ and S${}_2$, and scattered wave is observed at the point P.}
	\label{fig:DS}
\end{figure}
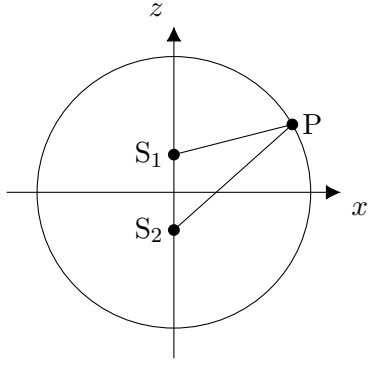

Let us
put two slits $S_{1},\: S_{2}$
on $(z,x) = (\pm l/2, 0)$,
and place
the center of a spherical screen of redius $L \gg l$
at the origin.
A wave with wavelength $\lambda_{\text{ds}}$
passes through the slits
from the negative
to the positive $x$-direction.
An observer P
is placed
at $(z,x)=(L\cos\theta', L\sin\theta')$ on the screen.
See Fig.~\ref{fig:DS}.

If the screen is
sufficiently large, {\it i.e.} $L \gg l$,
the difference of two optical paths
is given by
\begin{align}
	\overline{\text{S}_2\text{P}} - \overline{\text{S}_1\text{P}}
	\simeq
	l \cos\theta'.
\end{align}
Thus the condition that
the amplitude at $\text{P}$ vanishes
is
\begin{align}
	\label{eq:slit_zero_cond}
    \cos\theta'
	\simeq \frac{k'}{2} \frac{\lambda_{\text{ds}}}{l}
\end{align}
where $k'$ is an odd integer.
More precisely, supposing that a spherical wave is emitted from the two slits with identical phase and amplitude $A$, 
the wave amplitude on the sphere placed at the spatial infinity $L (\gg l)$ is
\begin{align}
\label{eq:DSamp}
    {\mathcal A}(\theta',\varphi')
    \simeq 2A \cos\left(\frac{\pi l}{\lambda_{\text{ds}}}\cos\theta'\right).
\end{align}
Zeros of this amplitude is given by \eqref{eq:slit_zero_cond}.
The result is independent of another spherical coordinate $\varphi'$ as the slits are 
separated along the $z$-axis in this case.

\begin{figure}[t]
	\centering
	\includegraphics[width=60mm]{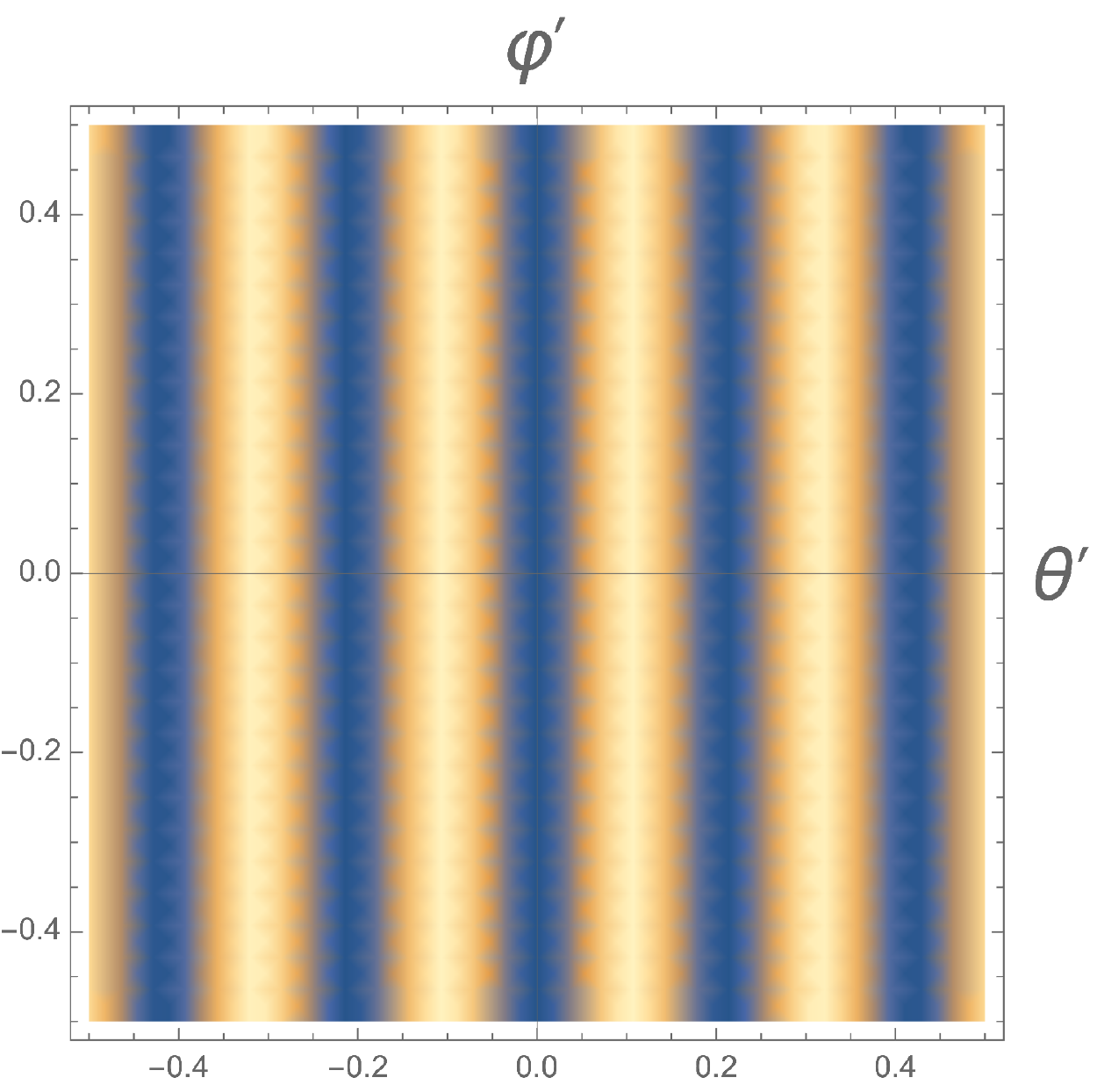}
	\caption{The amplitude of waves in the double slit experiment, for $\pi l/\lambda_{\text{ds}} = 15$.}
	\label{fig:DSamp}
\end{figure}

The wave amplitude itself does not give the image of the slit, as understood in Fig.~\ref{fig:DSamp}. 
Optical wave theory provides a method for the imaging. The popular method 
is just to put a lens of a finite size $d$ on the celestial sphere, and convert the wave amplitude with the frequency $\omega=2\pi/\lambda_{\text{ds}}$ into the image on a virtual screen
located at a focal point of the lens. The formula for converting the amplitude to the image is approximated just by a Fourier transformation for $L\gg l$
(see \cite{Hashimoto:2018okj,Hashimoto:2019jmw} for more details and brief reviews of the formula):
\begin{align}
\label{eq:imaging_formula}
    I(X,Y) = \frac{1}{(2d)^2\sin\theta_0'} \int_{\theta_0'-d}^{\theta_0'+d} d\theta' \int_{\varphi_0'-d/\sin\theta_0'}^{\varphi_0'+d/\sin\theta_0'} d\varphi' e^{i \omega ((\theta'-\theta_0') X + (\varphi'-\varphi_0') Y)} {\mathcal A}(\theta', \varphi').
\end{align}
Here the coordinate $(X,Y)$ is the one on the virtual screen placed behind the lens, so the image $I$ of the optical wave is produced on
that $XY$-plane. The location of the center of the lens is at $(\theta', \varphi')=(\theta_0', \phi_0')$, and a rectangular shape of the lens region is adopted for computational simplicity. 

The lens size $d$ needs to cover at least several zeros of the amplitude to find 
a sharp image, otherwise the image is blur and no structure can be reconstructed from the amplitude.
In the present case of the double-slit experiment, when one takes the limit $l\gg \lambda_{\text{ds}}$ the zeros of the amplitude are aligned densely,
so the lens size $d$ can be taken quite small compared to $2\pi$. It means 
that the curvature effect of the celestial sphere in the imaging can be ignored and the formula \eqref{eq:imaging_formula} is validated.

\begin{figure}[t]
	\centering
	\includegraphics[width=60mm]{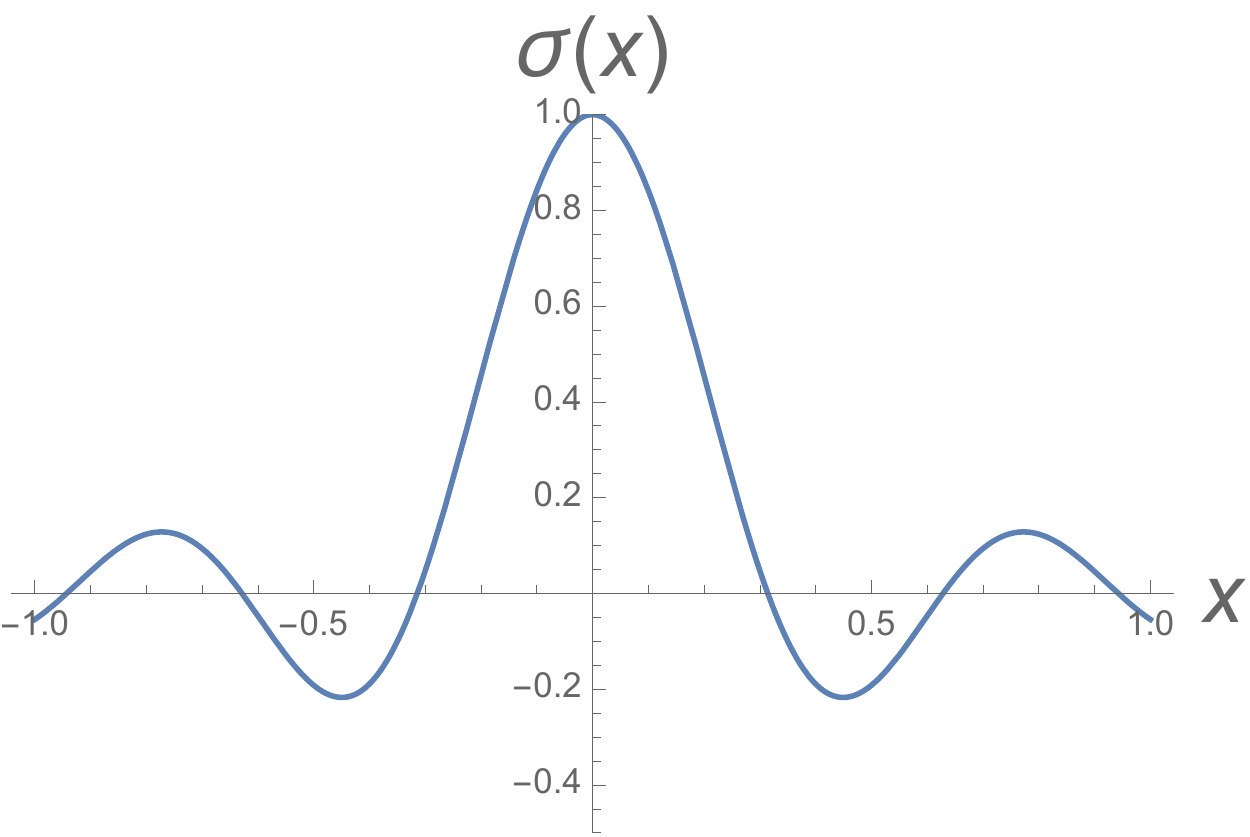}\hspace*{10mm}
	\includegraphics[width=60mm]{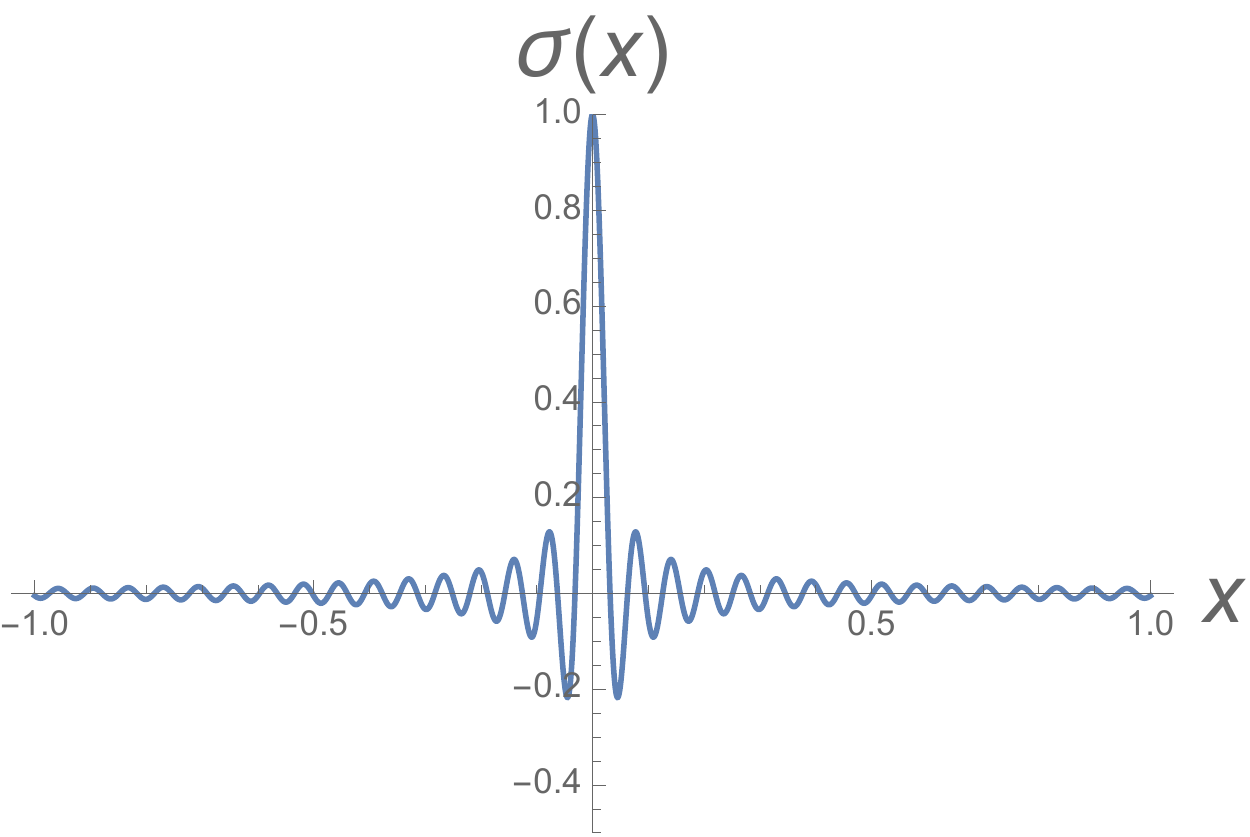}
	\caption{The lens resolution, seen in the behavior of the function $\sigma(x)$ \eqref{eq:sigma}, for $2 \pi d/\lambda_{\text{ds}}=10$ (left) and for $2 \pi d/\lambda_{\text{ds}}=100$ (right). For large $d/\lambda_{\text{ds}}$ (meaning that the wave length is small enough so that it detects many oscillations of the amplitude), this function is highly peaked at $x=0$, and the resolution is high.}
	\label{fig:sigma}
\end{figure}

Let us substitute the double-slit amplitude \eqref{eq:DSamp} into the imaging formula \eqref{eq:imaging_formula} and check if the image reproduces the shape (location) of the slits.
The result is
\begin{align}
\label{eq:DS_image}
    I(X,Y) = A(-1)^{\tilde{k}} \sigma(Y/\sin\theta_0')
    \left(
    \sigma\left(X-\frac{l\sin\theta_0'}{2}\right)
+
    \sigma\left(X+\frac{l\sin\theta_0'}{2}\right)
\right)
\end{align}
where we have defined
\begin{align}
\label{eq:sigma}
    \sigma(x)\equiv \frac{\sin\left(\frac{2\pi d }{\lambda_{\text{ds}}}x\right)}{\frac{2\pi d }{\lambda_{\text{ds}}}x}.
\end{align}
See Fig.~\ref{fig:sigma} for the meaning of this function $\sigma(x)$.
The calculation is straightforward but we have just assumed that the center of the location of the lens satisfies the condition
\begin{align}
    \frac{l}{\lambda_{\text{ds}}}\cos\theta_0' =\tilde{k}\in{\mathbf{Z}}
\end{align}
to simplify the result. 
It can be met easily because we can find such an integer in the limit $l \gg\lambda_{\text{ds}}$ which we took as described above.

\begin{figure}[t]
	\centering
	\includegraphics[width=40mm]{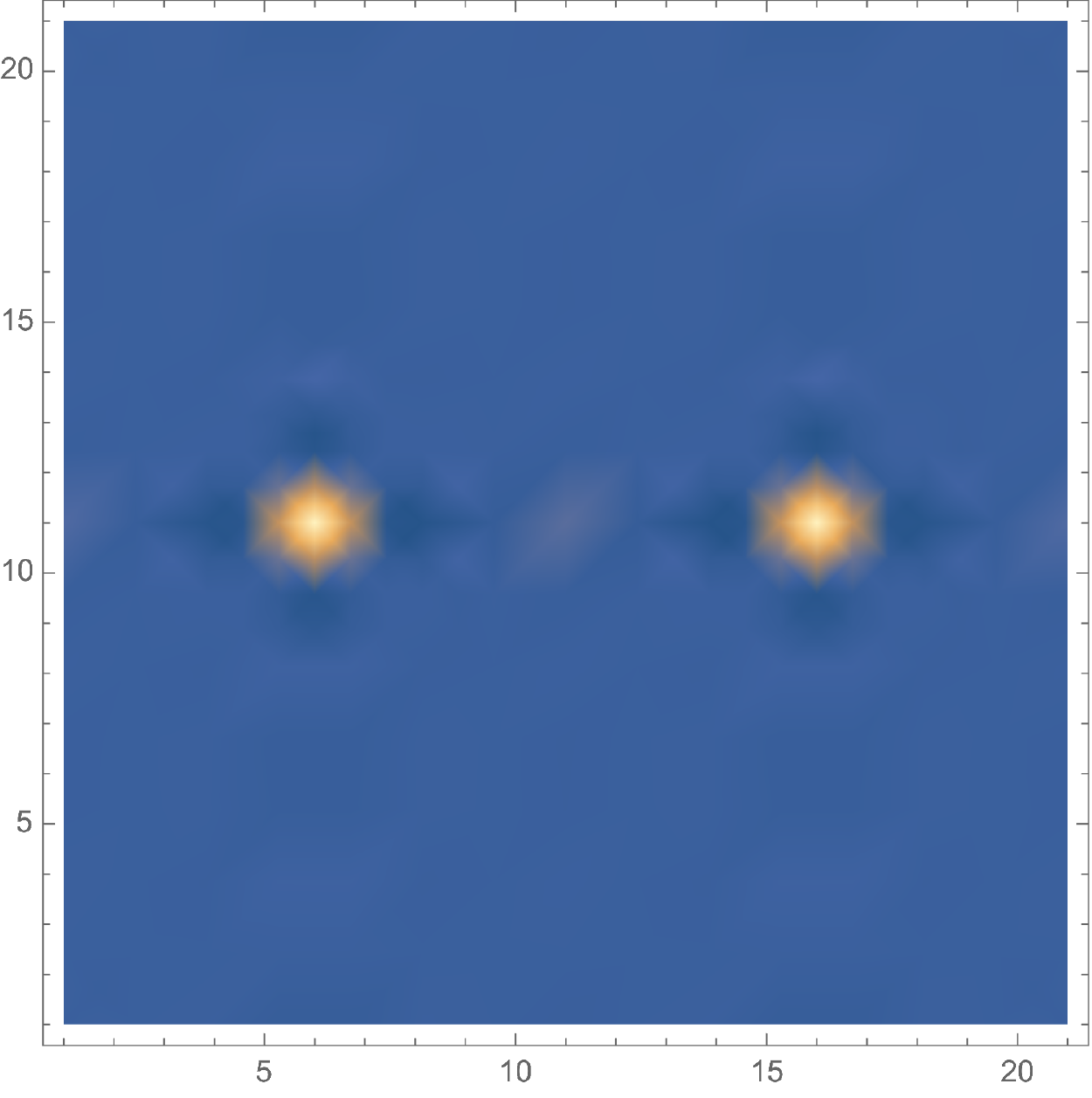}\hspace*{10mm}
	\includegraphics[width=40mm]{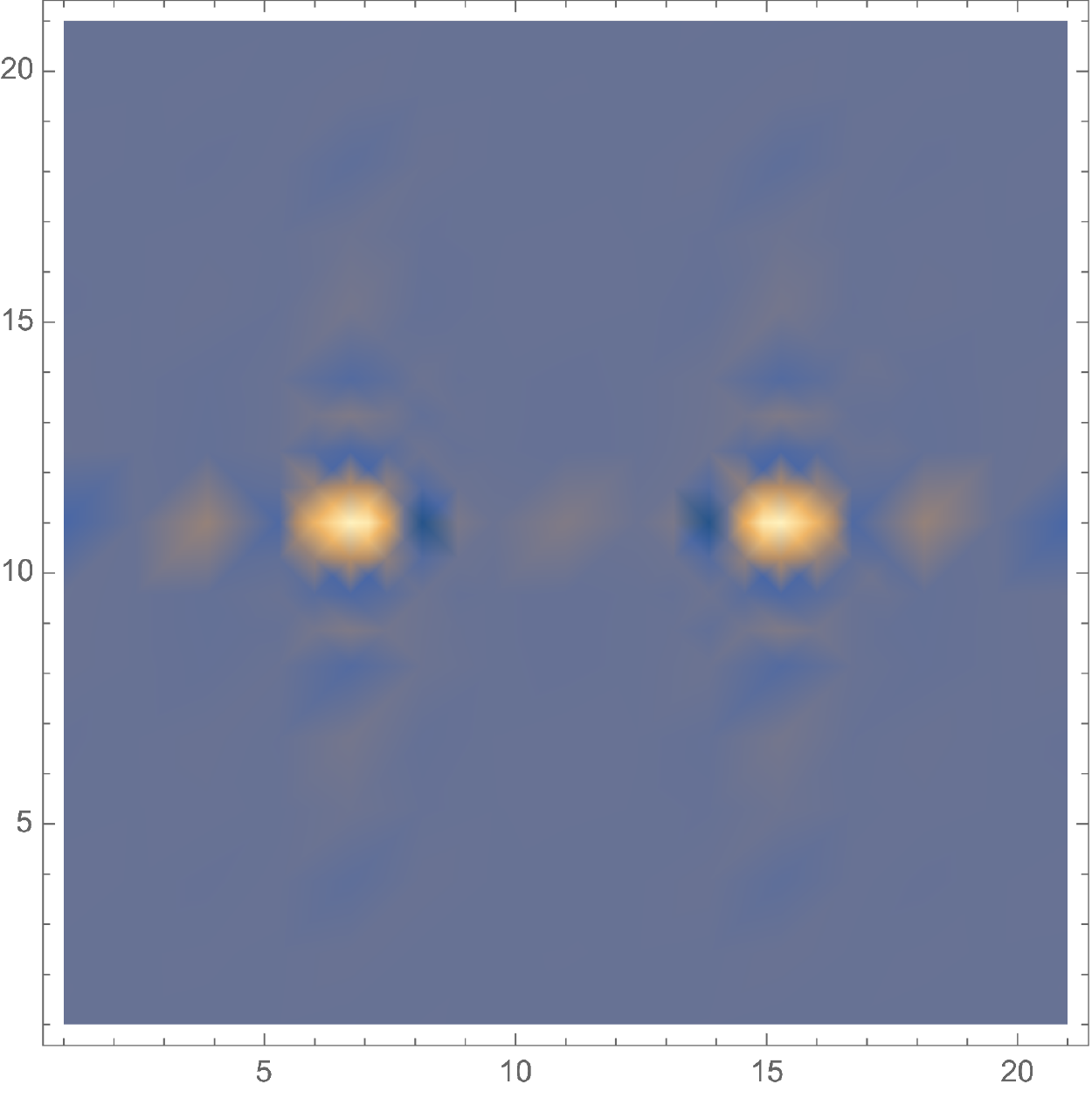}\hspace*{10mm}
	\includegraphics[width=40mm]{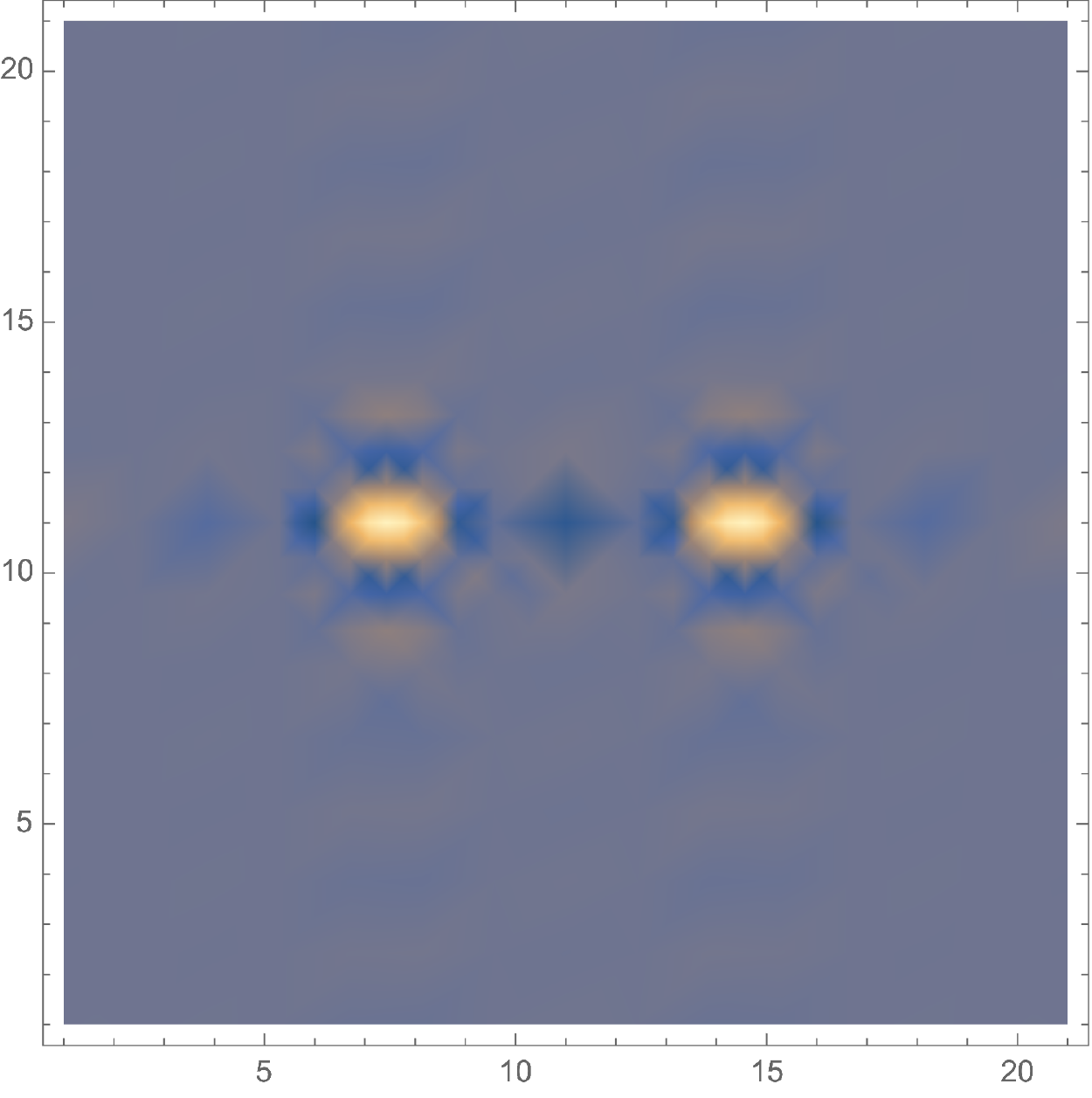}
	\caption{The $X$-$Y$ images of the double-slit experiment, reconstructed from the wave amplitudes at the spatial infinity. Left: seen from the head-on point $(\theta_0', \varphi_0')=(\pi/2,0)$ facing the aligned slit. Middle: $(\theta_0', \varphi_0')=(\pi/3,0)$. Right: $(\theta_0', \varphi_0')=(\pi/4,0)$. As the viewer shifts from the head-on point, the distance between the slits decreases by the function $\sin\theta_0'$ consistently. }
	\label{fig:DSimage}
\end{figure}

The density plots of the image function \eqref{eq:DS_image} are shown in Fig.~\ref{fig:DSimage}. One clearly finds the position of the double slits.
The parameters are chosen as: $2\pi d/\lambda_{\text{ds}}=30$ and $l=1$. The left figure is the image of the lens put at $(\theta_0', \varphi_0')=(\pi/2,0)$. The location $\theta_0'=\pi/2$ means that the lens is put at the right hand side in Fig.~\ref{fig:DS}, facing the double slits head-on, and the distance between the images of the two slits should be the largest. On the other hand, when the location of the viewer is shifted in the $z$ direction in Fig.~\ref{fig:DS}, the distance between the images of the slits should decrease. In fact, as seen in the middle and the right figures in Fig.~\ref{fig:DSimage}, for  $(\theta_0', \varphi_0')=(\pi/3,0)$ and for $(\theta_0', \varphi_0')=(\pi/4,0)$ respectively, one finds a consistent behavior.

In the next section, we apply this method for string scattering amplitudes, and see the images of the fundamental string.

\section{Images of string via Veneziano amplitude}
\label{sec:veneziano}

In this section,
we obtain the images of a fundamental string by using the Veneziano amplitude, and show that the fundamental string is a double-slit.

More specifically,
we point out that
the zeros of the Veneziano amplitude
match
the ones of the double-slit amplitude
described in the previous Sec.~\ref{sec:double-slit}.
Then
we reconstruct images of a string
by the Fourier transformation of the Veneziano amplitude.
The images show two columns (which are ``slits")
standing apart at a distance of a typical length of strings.

\subsection{Veneziano amplitude and its zeros}

First, we describe our notation of the Veneziano amplitude,
tachyon-tachyon to tachyon-tachyon scattering amplitude.

We label its four vertices with $1,2,2',1'$.
The momentum conservation law is
\begin{align}
	0 = p_1+p_2+p_2'+p_1'.
\end{align}
Since each vertex is a tachyon, we have the on-shell condition
\begin{align}
	p_1^2 = p_2^2 = p_2'^2 = p_1'^2 = -(-2).
\end{align}
Here, we conventionally chose $\alpha' = 1/2$.
In the following computations,
we will use the same unit.
The Mandelstam variables are defined as
\begin{align}
	&s = -(p_1+p_2)^2 = -(p_1'+p_2')^2, \\
	&t = -(p_1 + p_2')^2 = -(p_1'+p_2)^2, \\
	&u = -(p_1+p_1')^2=-(p_2+p_2')^2.
\end{align}
These satisfy an identity
\begin{align}
	\label{eq:veneziano_mandelstam_id}
	s+t+u = 4\cdot(-2).
\end{align}

The Veneziano amplitude \cite{Veneziano:1968yb} is given by
\begin{align}
	\label{eq:veneziano_amp}
	\mathcal{A}^{\text{Ven}}
	&= 2 \left[
		\mathcal{A}_{st}^{\text{Ven}} + \mathcal{A}_{tu}^{\text{Ven}} + \mathcal{A}_{us}^{\text{Ven}}
	\right],
\end{align}
where
\begin{align}
	\mathcal{A}_{xy}^{\text{Ven}}
	= \frac{ \Gamma(-\alpha(x))\Gamma(-\alpha(y)) }{ \Gamma(-\alpha(x)-\alpha(y)) }, \quad
	\alpha(x) = 1 + x/2.
\end{align}
The amplitude has $s$-channel poles
at $s=2n$ for integers $n \geq -1$.
Since we are interested in the images of a string formed at the scattering, we 
pull out one of the $s$-channel poles and look at the residue of the pole of the Veneziano amplitude.
Using the reflection formula of the Gamma function
and the identity \eqref{eq:veneziano_mandelstam_id},
the residue of the $s$-channel pole is
\begin{align}
	\label{eq:veneziano_amp_s}
	\tilde{\mathcal{A}}_{s}^{\text{Ven}}
	&= \lim_{s\rightarrow 2n}
	\frac{ \sin\pi(-1-s/2) }{ \pi }\:
	(\mathcal{A}_{st}^{\text{Ven}}+\mathcal{A}_{us}^{\text{Ven}}) \notag\\
	&= \frac{1}{\Gamma(2+n)} \left[
		\frac{ \Gamma(-1-t/2) }{ \Gamma(-2-n-t/2) }
		+ (t \leftrightarrow u)
	\right]
	= \frac{1}{\Gamma(2+n)} \left[
		\frac{ \Gamma(-1-t/2) }{ \Gamma(2+u/2) }
		+ (t \leftrightarrow u)
	\right].
\end{align}

To evaluate this amplitude, we denote 
the magnitude of the momentum of the in-coming tachyons is $p$. 
In the center-of-mass frame, 
the momenta can be generically parametrized as
\begin{align}
	\begin{array}{ll}
		p_1 = \mqty( \sqrt{p^2-2} \\ p\sin\theta \\ 0 \\ p\cos\theta ), & 	\hspace*{5mm}
		-p_1' = \mqty( \sqrt{p^2-2} \\ p\sin\theta'\cos\varphi' \\ p\sin\theta'\sin\varphi' \\ p\cos\theta' ), 		\vspace*{3mm}
\\
		p_2 = \mqty( \sqrt{p^2-2} \\ -p\sin\theta \\ 0 \\ -p\cos\theta ), &\hspace*{5mm}
		-p_2' = \mqty( \sqrt{p^2-2} \\ -p\sin\theta'\cos\varphi' \\ -p\sin\theta'\sin\varphi' \\ -p\cos\theta' ).
	\end{array}
\end{align}
The parameter $\theta$, 
which measures the angle of the incoming tachyon against the $z$-axis,
will be fixed 
later for the convenience in numerical computations.
Under this parametrization,
the Mandelstam variables are expressed as
\begin{align}
	s &= 4(p^2-2) = 2n, \\
	t &= -2p^2(1+\cos\theta\cos\theta'+\sin\theta\sin\theta'\cos\varphi' ), \\
	u &= -2p^2( 1-\cos\theta\cos\theta'-\sin\theta\sin\theta'\cos\varphi' ).
\end{align}

Let us look for zeros of the residue of the $s$-channel pole of 
the amplitude \eqref{eq:veneziano_amp_s}.
It is zero
if and only if
\begin{align}
	0
	&= \Gamma(-1-t/2)\Gamma(2+t/2) + (t \leftrightarrow u) \notag\\
	&= \frac{ \pi }{ \sin\pi(-1-t/2) } + (t \leftrightarrow u).
\end{align}
This condition is equivalent to
\begin{align}
	\pi(-1-t/2) + \pi(-1-u/2) &= 2\mathbf{Z} \pi, \\
	\text{or\;\;}
	\pi(-1-t/2) - \pi(-1-u/2) &= (2\mathbf{Z}-1) \pi.
\end{align}
If $n$ is even,
the first equation is satisfied
for any angles.
Then
the amplitude is trivially zero.
In the following discussion,
we assume that $n$ is odd.
The second equation reduces to
\begin{align}
	\label{eq:veneziano_amp_s_zero}
	\cos\theta\cos\theta'
	+\sin\theta\sin\theta'\cos\varphi'
	= \frac{2\mathbf{Z}-1}{n+4}.
\end{align}
Fixing $\theta=0$
makes clear
a comparison of this result
to the double-slit experiment in Sec.~\ref{sec:double-slit}.
In the present case
the zeros of the amplitude are at
\begin{align}
\cos\theta'= \frac{2k+n}{n+4},
\end{align}
for an integer $k \in \mathbf{Z}$.
Rewriting this as
\begin{align}
	\cos\theta'
	\simeq
	\frac{k'}{n+4}
	= \frac{k'}{2p^2} , 
\end{align}
where $k'$ is an odd integer, $k' \in 2\mathbf{Z}-1$,
we find that the location of the zeros is exactly the same as that of the double-slit experiment, \eqref{eq:slit_zero_cond},
with the replacement 
\begin{align}
    p^2 = \frac{l}{\lambda_{\text{ds}}} .
\end{align}
Here $p$ is the magnitude of the tachyon momentum in the Veneziano amplitude, while $l$ $(\lambda_{\text{ds}})$ is the slit-separation (wave length) in the double-slit experiment.

We have found here that the zeros of the Veneziano amplitude coincide completely with the double-slit experiment. In the next subsection, we look at the Veneziano amplitude in more detail and will find that indeed the imaging of the fundamental string results in a double-slit in the $n \to \infty$ limit.

%


\subsection{Imaging of a fundamental string}

Here first we present
images of a string
by straightforward numerical evaluations,
and after that, we analyze the images
analytically
in a certain limit to extract the structure.

The imaging formula described in Sec.~\ref{sec:double-slit} is
\begin{align}
\label{eq:Vene_image}
    I(X,Y) = \frac{1}{(2d)^2\sin\theta_0'} \int_{\theta_0'-d}^{\theta_0'+d} d\theta' \int_{\varphi_0'-d/\sin\theta_0'}^{\varphi_0'+d/\sin\theta_0'} d\varphi' e^{i p ((\theta'-\theta_0') X + (\varphi'-\varphi_0') Y)} \tilde{\mathcal A}_s^{\text{Ven}}(\theta', \varphi').
\end{align}
Here $d$ is the size of the lens whose center is put at $(\theta',\varphi')=(\theta_0', \varphi_0')$. The lens is applied to the wave signal $\tilde{\mathcal A}_s^{\text{Ven}}(\theta',\varphi')$ which reached the asymptotic infinity with the angle $(\theta',\varphi')$ in the 2-dimensional spherical coordinate.
The frequency of the wave, appearing in the formula \eqref{eq:imaging_formula}, is now given by the tachyon momentum $p$.
The wave amplitude is taken as 
the residue of the $s$-channel pole at $s=2n$, of the Veneziano amplitude 	\eqref{eq:veneziano_amp_s}.

\begin{figure}[t]
	\centering
	\includegraphics[width=150mm]{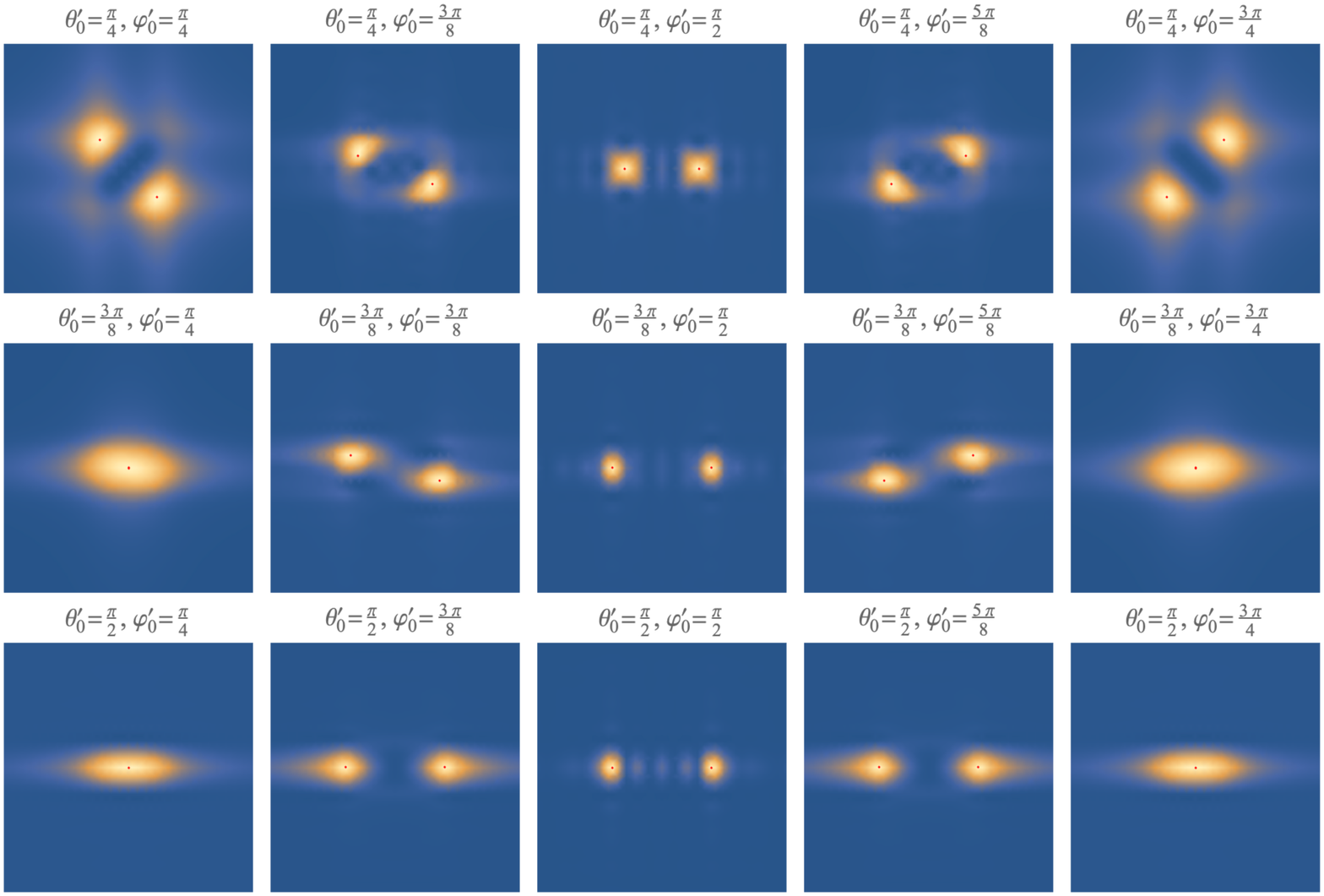}
	\caption{Images of a fundamental string, extracted and reconstructed from the $s$-channel pole of the Veneziano amplitude. The incoming angle is chosen as $\theta=\pi/2$. Various figures are for different choices of the viewer's location (lens location) specified by $(\theta_0',\varphi_0')$. Brightness of these density plots is normalized in each plot. The red points in the images are the brightest points.}
	\label{fig:Veneimage}
\end{figure}

\begin{figure}[t]
	\centering
	\includegraphics[width=150mm]{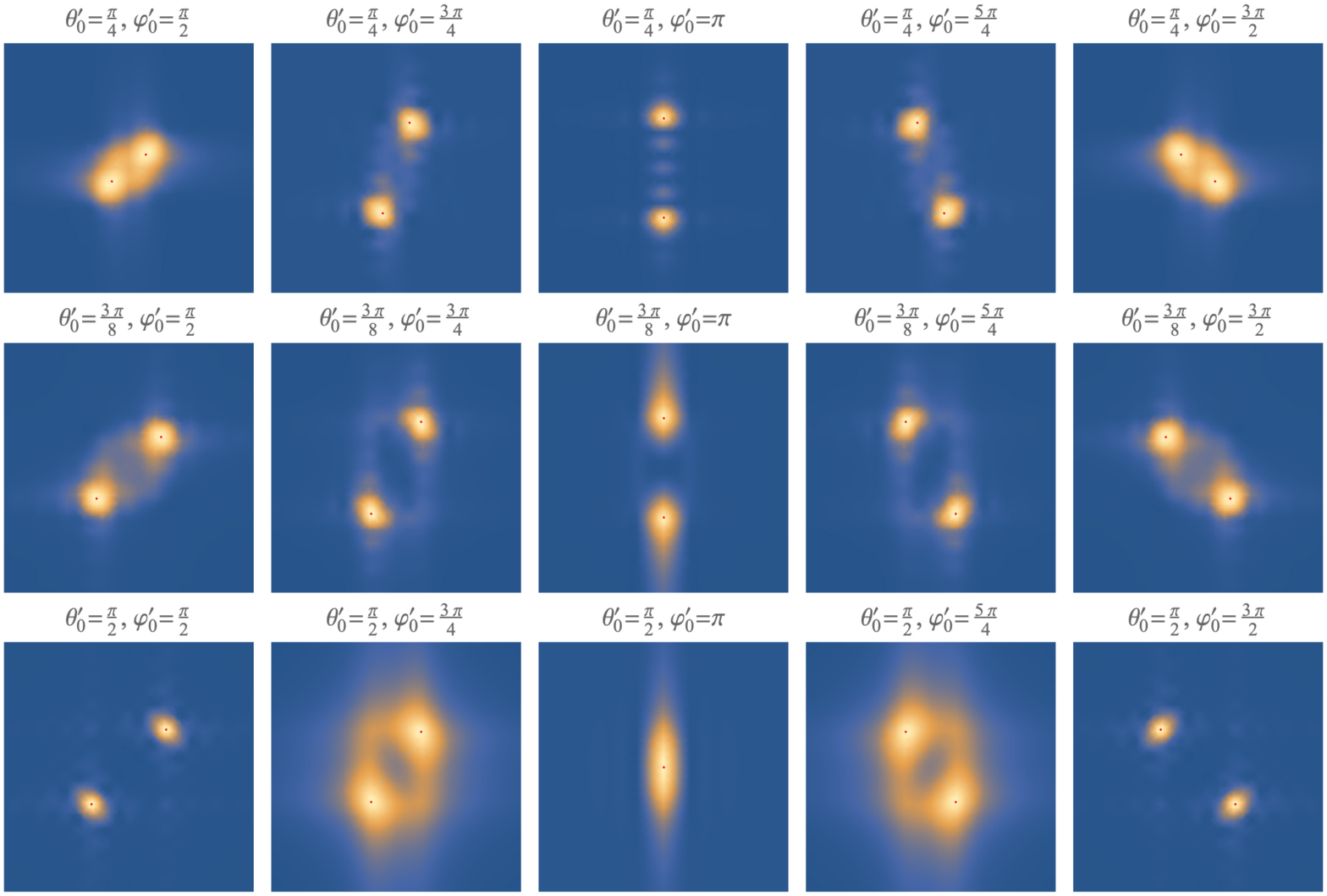}
	\caption{Images of a fundamental string, with $\theta=\pi/4$.}
	\label{fig:Veneimage45}
\end{figure}

A straightforward numerical evaluation of the imaging \eqref{eq:Vene_image} produces the images shown in Fig.~\ref{fig:Veneimage}
and Fig.~\ref{fig:Veneimage45}. 
Since the tachyon scattering is in the center-of-mass frame, we can fix $\theta$ without losing generality.  Fig.~\ref{fig:Veneimage} and Fig.~\ref{fig:Veneimage45}
are for $\theta=\pi/2$ and $\theta=\pi/4$, respectively, with $n=11$.
We find that in fact the images of the fundamental string are those of the double slits.

Analytic estimation of the image, which is useful to study the structure of the images, 
is possible. Let us investigate in particular the separation between the two slits. For simplicity let us take $\theta=0$ to eliminate the $\varphi'$-dependence of the amplitude \eqref{eq:veneziano_amp_s}. We find
\begin{align}
    \label{eq:amp_total}
    \tilde{\mathcal A}_{s}^{\text{Ven}}
    =\frac{1}{(n+1)!}
    \left[
    \frac{\Gamma\left(\frac{n}2 + 1 - \frac{n+4}{2}\cos\theta'\right)}{\Gamma\left(
    -\frac{n}{2}-\frac{n+4}{2}\cos\theta'
    \right)}
    +
    \frac{\Gamma\left(\frac{n}2 + 1 + \frac{n+4}{2}\cos\theta'\right)}{\Gamma\left(
    -\frac{n}{2}+\frac{n+4}{2}\cos\theta'
    \right)}
    \right]. 
\end{align}
To perform an analytic evaluation of the imaging formula for this amplitude, 
we need to find a simplified expression of the amplitude. A simplification occurs when we take a large $n$. Indeed, in the large $n$ limit the zeros are very close to each other, thus a small lenz is sufficient for the imaging. This means that we need only a local expression of \eqref{eq:amp_total} around a certain value of $\theta'$. 

Below, we obtain a simplified local expression of the amplitude around a zero specified by the integer 
$k$
given in \eqref{eq:veneziano_amp_s_zero}.
We present two expressions; one is for the region around the zero with $k=-(n+1)/2$, which means the expression for $\theta'\sim 0$ in the large $n$ limit, and the other is for a generic value of $k$.

\begin{figure}[t]
	\centering
	\includegraphics[width=80mm]{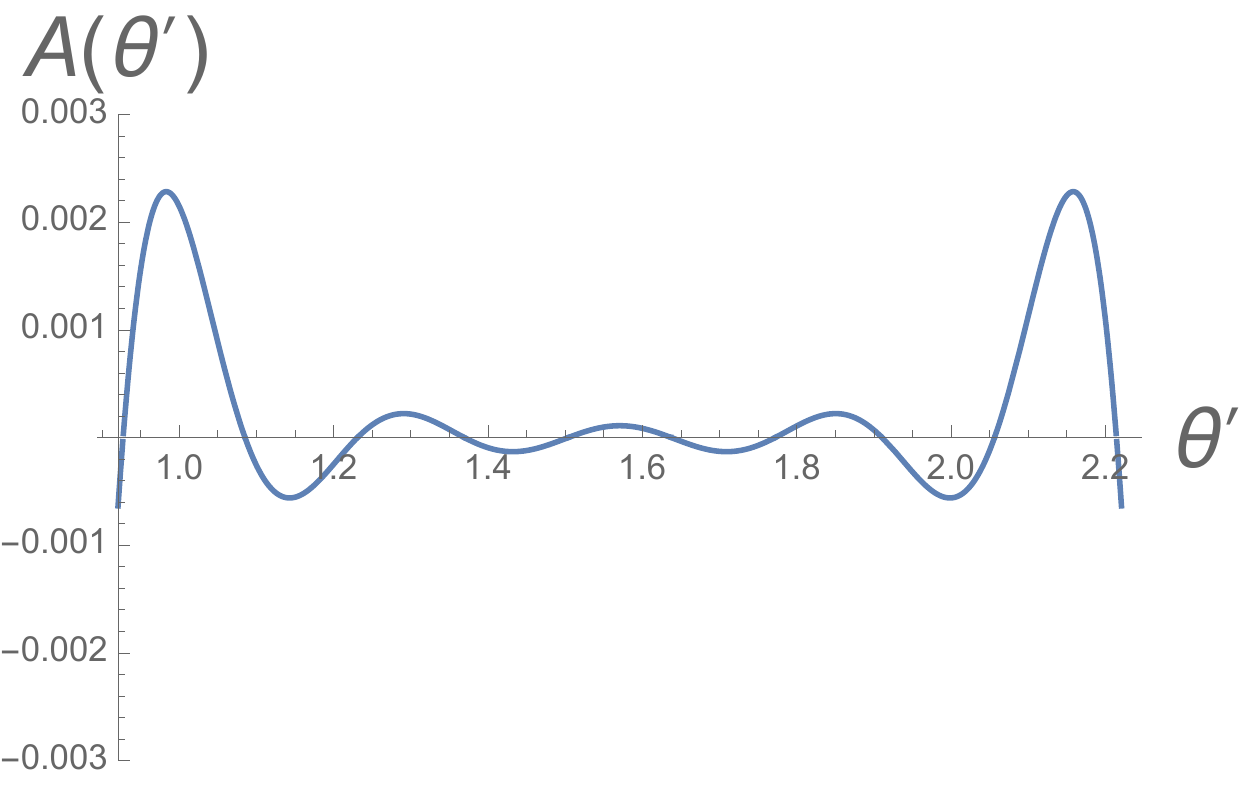}
	\caption{A plot of the amplitude \eqref{eq:amp_total} for $n=11$, magnified around $\theta'=\pi/2$.}
	\label{fig:amp_simple}
\end{figure}

First, notice that the amplitude \eqref{eq:amp_total} can be locally approximated by 
\begin{align}
\label{eq:amp_simp}
        \tilde{\mathcal A}_{s}^{\text{Ven}}
\sim f(\theta') \cos(c \theta'),
\end{align}
where $c$ is a constant dependent on the choice of $k$ of the central zero, and $f(\theta')$ is some smooth positive function giving the magnitude of the oscillation.
This is obvious if we look at 
the plot of the amplitude \eqref{eq:amp_total} given in Fig.~\ref{fig:amp_simple}. The amplitude has zeros at \eqref{eq:veneziano_amp_s_zero} and does not diverge, so it can be approximated by the form \eqref{eq:amp_simp} with a very simple function $f(\theta')$.

The derivation of the explicit form of $f(\theta')$ and the constant $c$ is simple. The latter can be fixed by the distribution of the location of the zeros given in \eqref{eq:veneziano_amp_s_zero}, and the magnitude function $f(\theta')$ can be fixed by evaluating the slope of the amplitude at those zeros. A straightforward calculation shows
\begin{align}
            \tilde{\mathcal A}_{s}^{\text{Ven}}(\theta')
\sim \pm
\left[{}\frac{2}{\pi \, {}_nC_\frac{n}{2}} 
\right] 
\cos\left(
\frac{n\pi}{2} 
\left(\theta'-\frac{\pi}{2}\right)\right)
\qquad (\theta' \sim \theta_0'=\pi/2)
\label{eq:amp_simp_90}
\end{align}
This expression tells us that the magnitude function $f(\theta')$ is constant around $\theta'=\pi/2$.
And for a generic $k$ (which means $\theta'\sim\theta_0'$ for generic $\theta_0'$), we find
\begin{align}
            \tilde{\mathcal A}_{s}^{\text{Ven}}(\theta')
\sim 
\pm
\left[\frac{2}{n\pi \, {}_nC_{\frac{n}{2}(1-\cos\theta_0')}}
\left(\frac{1-\cos\theta_0'}{1+\cos\theta_0'}\right)^{(n\sin\theta_0'/2)(\theta'-\theta_0')}
\right]
\cos\left(
\frac{n\pi\sin\theta_0'}{2} (\theta'-\theta_0')\right).
\label{eq:amp_simp_gene}
\end{align}
We can see that the magnitude function $f(\theta')$ is an exponentially growing function. 
As a consistency check, if one puts $\theta_0'=\pi/2$ in this expression \eqref{eq:amp_simp_gene}, it reduces to \eqref{eq:amp_simp_90}\footnote{Note that the difference factor $1/n$ is irrelevant as it comes out by just how the combinatoric factor ${}_nC_{\frac{n}{2}(1-\cos\theta_0')}$ is evaluated in the sub-leading order in the large $n$ expansion. In any case, the overall constant factor is not the issue to look at the structure of the images.}.

Let us evaluate the imaging formula \eqref{eq:Vene_image} with these large-$n$ approximated amplitudes \eqref{eq:amp_simp_90} and \eqref{eq:amp_simp_gene}. 
For the case $\theta'\sim\pi/2$, the expression \eqref{eq:amp_simp_90} is completely the same
as that of the double slit \eqref{eq:DSamp}, and we find
\begin{align}
    I_{\theta_0=\pi/2}(X,Y)  \propto \sigma(Y)
        \left[
    \sigma\left(X-\pi p\right)
+
    \sigma\left(X+\pi p\right)
\right]
\end{align}
where 
\begin{align}
    \sigma(x)\equiv \frac{\sin(pdx)}{pdx}. 
\end{align}
The function $\sigma$ is, as was described for the case of the double-slit experiment, is highly peaked at $x=0$ in the limit $p\to\infty$. Thus the bright spots in the image are 
localized at 
\begin{align}
    (X,Y)=(\pm\pi p,0).
\end{align}
This is in fact a double slit. In the limit $n \simeq 2p^2 \to \infty $, the
image consists of just two bright point-like spots, and the coincidence with the double-slit experiment is exact.

The distance $\Delta$ between the two bright points in the image is given by
\begin{align}
    \Delta = \pi\sqrt{2n}
\end{align}
in the string length unit $l_s=1/\sqrt{2}$ which we took throughout the paper.
The behavior $\Delta \sim \sqrt{n}\, l_s$ is consistent with the approximate length of a fundamental string at the excitation level $n$, as we will discuss in Sec.~\ref{sec:interpretation}.

Next, let us evaluate the images with the generic $\theta_0'$ using \eqref{eq:amp_simp_gene}. The result is
\begin{align}
    I_{\theta_0=\theta_0'}(X,Y)  
    \, \propto \, \sigma\!\left(\frac{Y}{\sin\theta_0'}\right)
        \left[
    \sigma\!\left(X-\frac{\Delta(\theta_0')}{2}+ i \tilde{X}_0 \right)
+
    \sigma\!\left(X+\frac{\Delta(\theta_0')}{2}+ i \tilde{X}_0 \right)
\right]
\end{align}
where 
\begin{align}
        \Delta(\theta_0') \equiv  \pi\sqrt{2n}\sin\theta_0', \quad 
        \tilde{X}_0 \equiv\sqrt{\frac{n}{2}}(\sin\theta_0')\log\frac{1+\cos\theta_0'}{1-\cos\theta_0'}.
\end{align}
This means that there are two slits which are located, in the complexified coordinates, 
\begin{align}
    (X,Y) = \left(\pm \frac{\Delta(\theta_0')}{2}+ i \tilde{X}_0, 0\right).
\end{align}
The imaginary part contributes to make the peak of $\sigma(x)$ smoother, as the pole deviates from the real axis.
The real part gives a physical consistency with the spatial location of the double-slit: the distance between the bright spots
in the image is $\Delta(\theta_0')$.
The factor $\sin\theta_0'$ in this $\Delta(\theta_0')$ is expected from the spatial consistency since the position of the viewer is shifted from the equator and thus
the two slits should be seen at an angle, exactly giving the factor $\sin\theta_0'$. Thus we conclude that the images seen at angle are spatially consistent with each other, which is a nontrivial check for the image slits to be present in the real space.

\section{Images of highly excited string}
\label{sec:HES}

In this section,
we show that
the slit-like behavior of a string
is
not limited to the Veneziano amplitude.
As another clean example,
we study
a highly excited string (HES) decaying to two tachyons.

The amplitude formula
for a HES decaying to two tachyons
was computed in \cite{Gross:2021gsj,Rosenhaus:2021xhm}.
We will briefly review their results
but with slight modifications.
Later,
we
list the location of the amplitude zeros,
and point out that
its amplitude zeros
also match
the ones of the double-slit
in Sec.~\ref{sec:double-slit}.
Then we reconstruct
images of a HES
by the Fourier transformation of the amplitude.
The images
show multi-slits,
which are understood as convolutions of double-slits.

\subsection{Amplitude and its zeros}

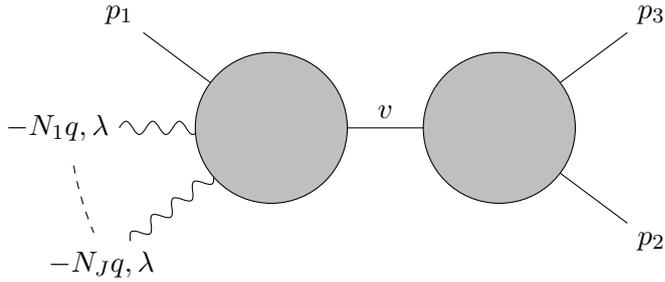
\begin{figure}[t]
	\centering
	\hspace*{-18mm}
	\begin{tikzpicture}
		\node[anchor=south] (c) at (0,0) {$v$};
		\node (dl) at (-1.5,0) {};
		\node (dr) at (1.5,0) {};
		
		\node (t1) at (-3.5,1.5) {$p_1$};
		\node[anchor=east] (p1) at (-3.5,0) {$-N_1q,\lambda$};
		\node[anchor=north east] (pJ) at (-2.9,-1.5) {$-N_Jq,\lambda$};

		\node (t3) at (3.5,1.5) {$p_3$};
		\node (t2) at (3.5,-1.5) {$p_2$};

		\draw (dl) -- (dr);
		\draw (t1) -- (dl);
		\draw (t2) -- (dr);
		\draw (t3) -- (dr);
		\draw[snake=coil,segment aspect=0] (p1) -- (dl);
		\draw[snake=coil,segment aspect=0] (pJ) -- (dl);
		
		\draw[fill=lightgray] (dl) circle (1);
		\draw[fill=lightgray] (dr) circle (1);
		
		\draw[dashed] (-4.1,-0.5) arc (190:203:4.1);
	\end{tikzpicture}	
	\caption{
		String amplitude
		of three tachyons and $J$ photons.
		Picking out a pole at the internal line labeled with $v$
		leads to
		the amplitude of
		a HES decaying into two tachyons.
	}
	\label{fig:hes_to_2tachyon}
\end{figure}

Following \cite{Rosenhaus:2021xhm},
we start with a string scattering amplitude for
\begin{center}
	tachyon, $J$ photons
	$\rightarrow$
	2 tachyons
\end{center}
as depicted in Fig.~\ref{fig:hes_to_2tachyon}.
After computing the amplitude,
we pick out a pole
which corresponds to an intermediate HES state.
Momenta of one incoming tachyon and two outgoing tachyons are denoted by $p_1$, $p_2$ and $p_3$. Momenta of $J$ photons are chosen to be proportional to a null vector, $q$, 
and are given by $-N_a q$, where $N_a$ are integers and $a=1,\cdots,J$. 
The momentum conservation law is
\begin{align}
	0 = (p_1-Nq)+p_2+p_3
\end{align}
where
\begin{align}
	N=\sum_{a=1}^{J} N_a, \quad
	(N_a\geq 1)
\end{align}
When a pole of the intermediate state is picked out, 
$N$ and $J$
will be interpreted
as the excitation level and the angular momentum
of the HES, respectively.
The on-shell conditions
for the tachyon momenta are
\begin{align}
	p_1^2 = p_2^2 = p_3^2 = -(-2).
\end{align}
For simplicity, polarizations of photons are taken to be equal to each other 
and denoted by $\lambda$. 
Then, the conditions for the photon momenta and polarizations are given by
\begin{align}
	q^2 = 0, \quad
	\lambda^2 = 0, \quad
	q\cdot \lambda = 0.
\end{align}
For simplicity,
we assume that
\begin{align}
	(p_2+p_3)\cdot\lambda = 0.
\end{align}
This condition is satisfied,
for example,
in the center-of-mass frame.
Then
the amplitude is factorized as 
\begin{align}
	\mathcal{A}^{\text{HES}}
	&= 
	\prod_{a=1}^J \mathcal A_a, 
\\
    \mathcal A_a
	&= \frac{ (-p_2\cdot\lambda)-(-p_3\cdot\lambda) }{2}
	\notag\\
	&\quad \times
			\left[
				\frac{ \Gamma(-\alpha_1^{(a)} +1)\Gamma(-\alpha_2^{(a)}) }
				{ \Gamma(-\alpha_1^{(a)}-\alpha_2^{(a)}+1) }
				-\frac{ \Gamma(-\alpha_2^{(a)})\Gamma(-\alpha_3^{(a)}) }
				{ \Gamma(-\alpha_2^{(a)}-\alpha_3^{(a)}) }
				+\frac{ \Gamma(-\alpha_3^{(a)})\Gamma(-\alpha_1^{(a)}+1) }
				{ \Gamma(-\alpha_3^{(a)}-\alpha_1^{(a)}+1) }
			\right],
\end{align}
where
\begin{align}
	\alpha_i^{(a)} = N_a p_i\cdot q.
\end{align}
Each factor of the amplitude, $\mathcal A_a$ 
has poles when $1 - \alpha_1^{(a)}$ is a negative integer. 
Poles at 
\begin{align}
    \alpha_1^{(a)} \sim N_a , 
\end{align}
are located at the same position in the momentum space, 
\begin{align}
	v = -(p_1-Nq)^2
	\sim 2(N-1) .
\end{align}
Picking out this pole, 
we have a HES
at an intermediate state.
By taking the residues of $\mathcal A_a$, we obtain 
the decay of the HES with the mass
\begin{align}
	M^2 = 2(N-1),
\end{align}
and the total angular momentum $J$.
Using an identity
$\alpha_1^{(a)}+\alpha_2^{(a)}+\alpha_3^{(a)}=0$,
and the condition of the pole, 
$\alpha_1^{(a)}\sim N_a$,
we find
\begin{align}
	\label{eq:hes_2tachyon_amp_v}
	\tilde{\mathcal{A}}_v^{\text{HES}}
	&= \lim_{v\rightarrow2(N-1)}
		\left( \prod_{a=1}^{J}  \frac{ \sin\pi\alpha_1^{(a)} }{ \pi } \right)
		\mathcal{A}^{\text{HES}} \notag\\
	&= \prod_{a=1}^{J}
			\frac{ (-p_2\cdot\lambda)-(-p_3\cdot\lambda) }{2}
			\left[
				\frac{\Gamma(-\alpha_2^{(a)})}{ \Gamma(N_a)\Gamma(-\alpha_2^{(a)}-N_a+1)}
				+
				(2\leftrightarrow 3)
			\right] \notag\\
	&= \prod_{a=1}^{J}
			\frac{ (-p_2\cdot\lambda)-(-p_3\cdot\lambda) }{2}
			\left[
				\frac{\Gamma(-\alpha_2^{(a)})}{ \Gamma(N_a)\Gamma(\alpha_3^{(a)}+1)}
				+ (2\leftrightarrow 3)
			\right]
\end{align}
Here,
the first term
is a contribution studied in \cite{Rosenhaus:2021xhm}.


In the center-of-mass frame,
we take the following parametrization of 
the momenta, 
\begin{align}
		&-q = \frac{1}{\sqrt{2N-2}} \mqty( 1 \\ 0 \\ 0 \\ -1 ), &
		\lambda &= \frac{1}{\sqrt{2}} \mqty( 0 \\ 1 \\ i \\ 0 ),
        \\
		&p_1-Nq = \mqty( \sqrt{2N-2} \\ 0 \\ 0 \\ 0 ) , &
		-p_2 &= \mqty( \sqrt{2N-2}/2 \\ \sqrt{2N+6}/2\: \sin\theta'  \\ 0 \\ \sqrt{2N+6}/2\: \cos\theta'),
&		-p_3 &= \mqty( \sqrt{2N-2}/2 \\ -\sqrt{2N+6}/2\: \sin\theta'  \\ 0 \\ -\sqrt{2N+6}/2\: \cos\theta').
\end{align}
Under this parametrization,
the variables $\alpha_i^{(a)}$ are expressed as
\begin{align}
	&\alpha_1^{(a)} 
	= 
	N_a, 
	\\
	&\alpha_2^{(a)} 
	= 
	-\frac{N_a}{2}\left( 1+\frac{\sqrt{2N+6}}{\sqrt{2N-2}}\cos\theta' \right), 
	\\
	&\alpha_3^{(a)} 
	= 
	-\frac{N_a}{2}\left( 1-\frac{\sqrt{2N+6}}{\sqrt{2N-2}}\cos\theta' \right).
\end{align}

Let us look for zeros of the amplitude \eqref{eq:hes_2tachyon_amp_v}.
It is zero
if and only if
\begin{align}
	&0 = (-p_2\cdot\lambda) = -(-p_3\cdot\lambda), \\
	\text{or\;}\hspace*{5mm}
	&0 = \Gamma(-\alpha_2^{(a)})\Gamma(\alpha_2^{(a)}+1) + (2\leftrightarrow3)
\end{align}
for some $1\leq a \leq J$.
Calculations
similar to those in Sec.~\ref{sec:veneziano}
show that
the amplitude is trivially zero
if some $N_a$ is even.
In the following discussion,
we assume that
$N_a$'s are odd for any $1\leq a \leq J$.
Then
we find that
the zeros of the amplitude
are at
\begin{align}
	\frac{ \sqrt{2N+6} }{ \sqrt{2N-2} }
	\cos\theta'
	= \frac{ 2\mathbf{Z}-1 }{ N_a }.
\end{align}
In the high excitation limit $N_a \rightarrow \infty$,
it is written as
\begin{align}
	\cos\theta'
	\simeq
	\frac{ k' }{ N_a },
\end{align}
where $k'$ is an odd integer, $k'\in 2\mathbf{Z}-1$.
We find that the location of the zeros is exactly the same as that of the double-slit experiment, \eqref{eq:slit_zero_cond},
with the replacement 
\begin{align}
	\frac{N_a}{2} = \frac{l}{\lambda_{\text{ds}}} .
\end{align}
Here $N_a$ is an excitation level of the $a$-th mode
such that $N = \sum_a N_a$,
while $l$ $(\lambda_{\text{ds}})$ is the slit separation (wave length) in the double-slit experiment.

\subsection{Imaging of a highly excited string}

Here first we present
images of a HES
by straightforward numerical evaluations,
and then understand the images analytically in a certain limit.

\begin{figure}[t]
	\centering
	\includegraphics[width=70mm]{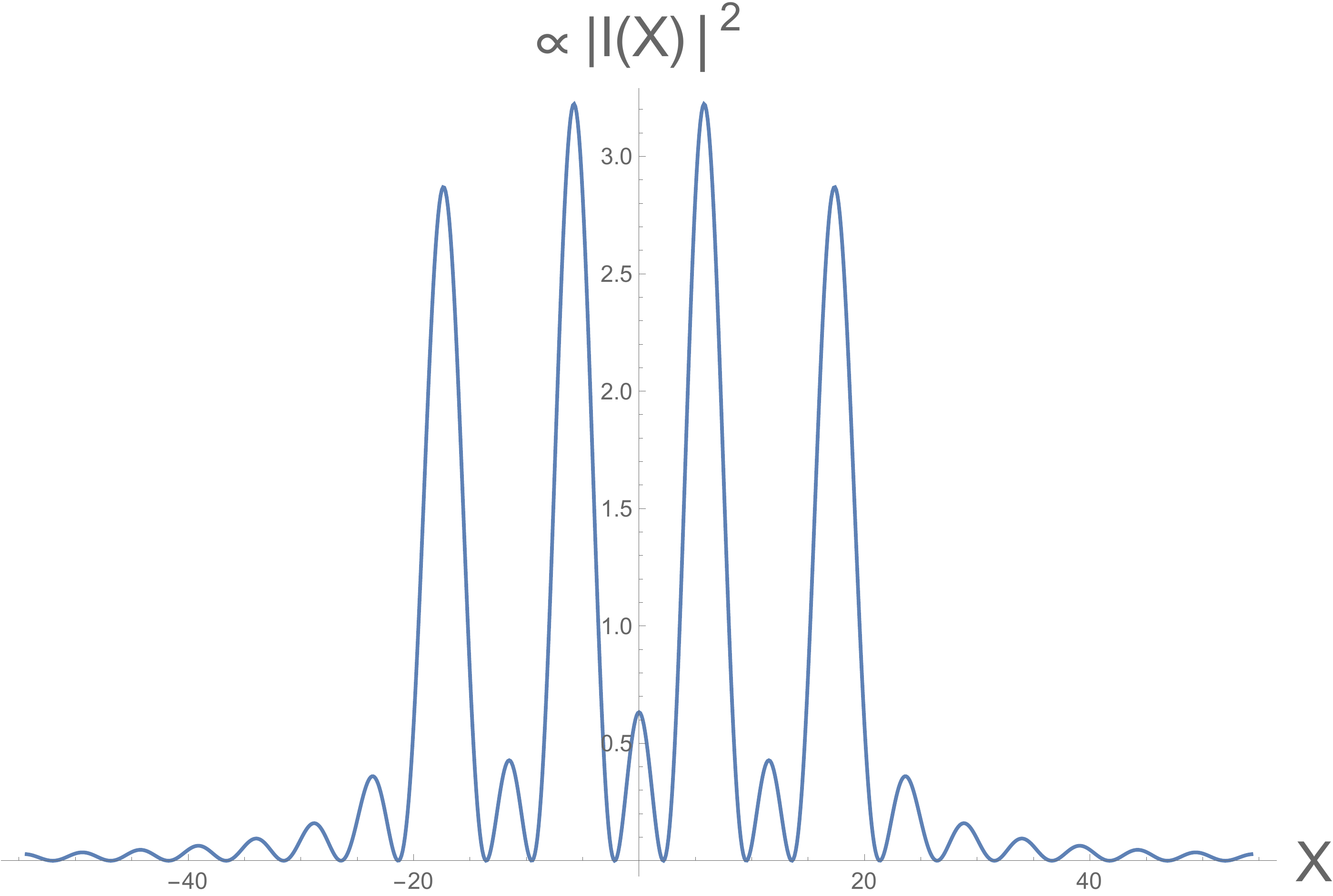}
	\hspace*{5mm}
	\includegraphics[width=70mm]{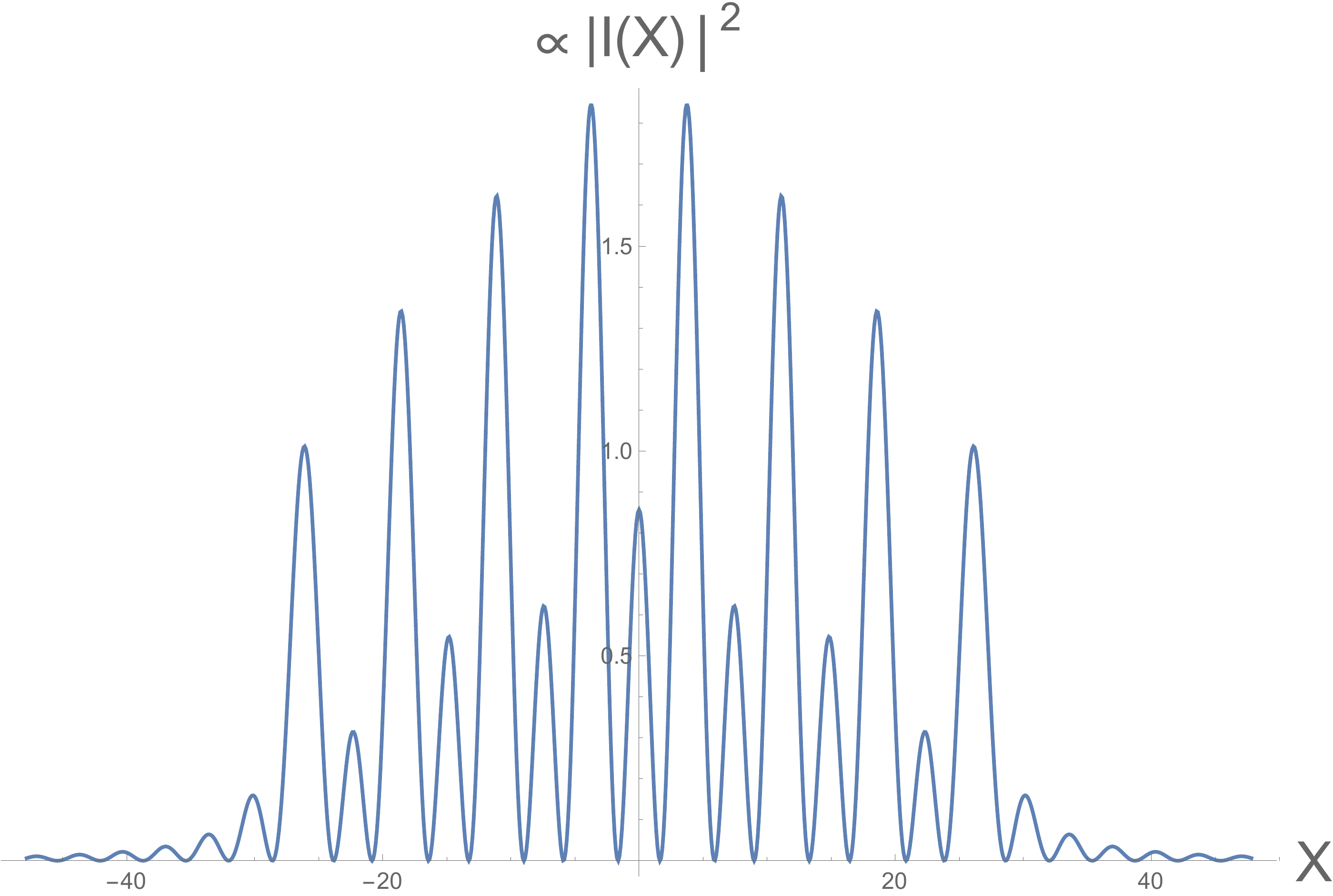}
	\caption{
		Images of a highly excited string,
		extracted and reconstructed
		from a HES intermediate state pole
		of the tachyon-photon amplitude.
        The size of the imaging lens
        is chosen as
        $d=0.11$.
		The left panel,
		for a division $N=N_1+N_2=19+39$,
		shows a $4$-slit,
		while the right panel,
		for $N=N_1+N_2+N_3=19+39+79$,
		shows an $8$-slit.
	}
	\label{fig:image_hes_2t}
\end{figure}

Since
the amplitude
depends only on the outgoing angle $\theta'$,
we restrict ourselves to
imaging in this single direction.
We use the following imaging formula
\begin{align}
	\label{eq:HES_image}
	I(X) = \frac{1}{(2d)^2\sin\theta_0'} \int_{\theta_0'-d}^{\theta_0'+d} d\theta'
	e^{i p (\theta'-\theta_0') X} \tilde{\mathcal A}_v^{\text{HES}}(\theta').
\end{align}
Here $p$ is now
the magnitude of the outgoing tachyon momentum
$p = \sqrt{2N+6}/2$.
A straightforward numerical evaluation of the imaging formula \eqref{eq:HES_image}
produces
the images shown in Fig.~\ref{fig:image_hes_2t}.
Both of the images
are for $\theta_0'=\pi/2$,
but with different numbers of excitations of $N = \sum_{a=1}^{J} N_a$.
We find that
the images of a HES
are those of multi-slits.

Such multi-slit structures
can simply be understood
by using the results in the Veneziano amplitude.
When $N_a \gg 1$,
the amplitude \eqref{eq:hes_2tachyon_amp_v} reduces to a simple form
\begin{align}
	\tilde{A}_v^{\text{HES}}
	\simeq
	\prod_{a=1}^{J} \sin\theta'
	\left[
		\frac{ \Gamma\left( +\frac{N_a}{2}(1+\cos\theta') \right) }
		{ \Gamma\left( -\frac{N_a}{2}(1-\cos\theta') \right) }
		+
		\frac{ \Gamma\left( +\frac{N_a}{2}(1-\cos\theta') \right) }
		{ \Gamma\left( -\frac{N_a}{2}(1+\cos\theta') \right) }
	\right]
\end{align}
up to a constant.
Remark that
each factor inside $\prod_{a=1}^{J}$
coincides with
the Veneziano amplitude at $\theta=0$ \eqref{eq:amp_total}
with the replacement $n \rightarrow N_a$,
\begin{align}
	\tilde{\mathcal{A}}_v^{\text{HES}}
	\simeq
	( \sin\theta' )^J
	\prod_{a=1}^{J}
	\left. \tilde{\mathcal{A}}_s^{\text{Ven}} \right|_{n\rightarrow N_a}.
\end{align}
Thus
the reconstructed image of a HES,
the Fourier transformation of $\tilde{\mathcal{A}}_v^{\text{HES}}$,
is the convolution
of the images of the Veneziano amplitudes with $n\rightarrow N_a$ as,
\begin{align}
	I^{\text{HES}}
	\simeq
	(\sin\theta')^J
	\ast \left. I^{\text{Ven}} \right|_{n\rightarrow N_1}
	\ast \cdots
	\ast  \left. I^{\text{Ven}} \right|_{n\rightarrow N_J}.
\end{align}
Recall that
the image of the Veneziano amplitude $I^{\text{Ven}}$
is a double-slit with a slit separation $l/\lambda_{\text{ds}} \simeq n/2$.
With replacements
$n\rightarrow N_a, \lambda_{\text{ds}} \rightarrow 1/\sqrt{N}$,
the image of a HES
is the convolution of double-slits
with separations $\Delta_a \sim N_a/\sqrt{N}$.
Then
we can conclude that
the image of a HES
is a $2^J$-slit.
Indeed
it can be easily demonstrated that
the convolution of double-slits with different slit separations
is a multi-slit.
Let
\begin{align}
	I_a(X)
	=
	\delta \left( X-\frac{\Delta_a}{2} \right)
	+ \delta \left( X+\frac{\Delta_a}{2} \right)
\end{align}
be an ideal image of a double-slit
with the slit separation $\Delta_a$.
Then
its convolution
\begin{align}
	I_a \ast I_b (X)
	&= \int_{-\infty}^{\infty} \dd{X'} \: I_a(X') I_b(X-X') \notag\\
	&= \sum_{s_1,s_2=\pm 1}
	\delta \left( X+ s_1\frac{\Delta_a}{2} + s_2\frac{\Delta_b}{2} \right)
\end{align}
is a $2^2$-slit.
Performing similar operations recursively,
we can show that
$J$-times convolution
of the double-slit with different slit-separations
is a $2^J$-slit as\footnote{Particularly when
$\Delta_a / \Delta_1 \simeq 2^a$,
the slits are equally separated.}
\begin{align}
	I_1 \ast \cdots \ast I_J (X)
	&= \sum_{s_a=\pm 1} \delta \left( X + s_1\Delta_1 \cdots + s_J\Delta_J \right).
\end{align}

We conclude that the image of the highly excited string reconstructed from its decay amplitude is a multi-slit.

\section{Interpretation}
\label{sec:interpretation}

In this section,
we present several observations
from the results obtained above,
and provide their possible interpretations.

Firstly,
we point out that
the slit-like appearance of scattering amplitude images
is an inherent feature of strings.
Recall, for example, the Veneziano amplitude.
At the zeros of the amplitude \eqref{eq:veneziano_amp_s_zero},
each term of \eqref{eq:veneziano_amp_s}
becomes zero simultaneously.
There is no non-trivial cancellation between the two terms.
In other words,
the zeros of the amplitude
originate in
the gamma functions in the denominator.
These gamma functions
are also the source of poles
which correspond to
intermediate excited string states.
Thus
the existence of a series of zeros \eqref{eq:veneziano_amp_s_zero}
reflects
the existence of a series of intermediate states in string theory.
Remember that, contrary to string theory,
perturbative quantum field theories
have only a few intermediate states,
and scatterings
occur at a point.

Secondly,
it is remarkable that
the order of slit separtions
agrees with that of the total string length.
Recall that
the slit separations
in the Veneziano amplitude
and in the HES scattering amplitude (with $J=1$)
are
\begin{align}
    \Delta^{\text{Ven}} \sim \sqrt{n}\: l_s, \quad
    \Delta^{\text{HES}} \sim \sqrt{N}\: l_s,
\end{align}
respectively.
Here
$n$ is roughly the square of the invariant mass,
and $N$ is roughly the square of the mass of a HES.
On the other hand,
the total length of a string with a mass $M$
is roughly estimated as 
\begin{align}
    l_{\text{total}} \sim \frac{1}{\alpha'} M \sim \sqrt{N}\: l_s
\end{align}
since a string has a constant tension $\alpha'$.
Thus, the slit separation and string length coincide with each other, 
meaning that the string may extend between the slits.

\begin{figure}[t]
	\centering
	\begin{tikzpicture}
		\draw[->] (-2.2,0) -- (2.2,0) node[anchor=north west] {$x$};
		\draw[->] (0,-2.2) -- (0,2.2) node[anchor=south east] {$z$};
		
		\draw (0,0) circle (1.8);
		\filldraw[black] (0,0.5) circle (2pt) node[anchor=east]{$\text{S}_1$};
		\filldraw[black] (0,-0.5) circle (2pt) node[anchor=east]{$\text{S}_2$};
		\filldraw[black] (-0.7,0) circle (2pt) node[anchor=north east]{$\text{S}_3$};
		\filldraw[black] ({1.8*cos(deg(pi/6))},{1.8*sin(deg(pi/6))}) circle (2pt) node[anchor=west]
		{$\text{P}$};
		
		\draw (0,0.5) -- ({1.8*cos(deg(pi/6))},{1.8*sin(deg(pi/6))});
		\draw (0,-0.5) -- ({1.8*cos(deg(pi/6))},{1.8*sin(deg(pi/6))});
		\draw (-0.7,0) -- ({1.8*cos(deg(pi/6))},{1.8*sin(deg(pi/6))});
	\end{tikzpicture}
	\caption{Three slits are put at S${}_1$, S${}_2$, S${}_3$, and scattered wave is observed at the point P.}
	\label{fig:DS_s3}
\end{figure}
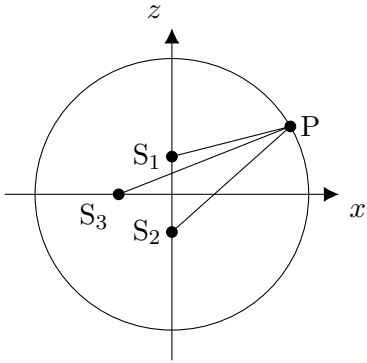

Thirdly,
we stress that
slits can only show up
on a straight line
both in the Veneziano amplitude
and the HES amplitude.
Technically, this is because
the $\theta'$ dependence of their amplitude zeros
is only in the form of $\pm \cos\theta'$.
In fact, other types of $\theta'$ dependence
with non-trivial phases $\cos(\theta'-\vartheta')$
are required
for the slit
to show up
away from the straight line.
This can be easily demonstrated as follows.
Suppose that
we have three slits $S_1,S_2,S_3$
which are not on a straight line
as in Fig.~\ref{fig:DS_s3}.
If the location of $S_3$ is
$(z,x)=(0,-l'/2)$,
the optical path lengthes are
\begin{align}
    \overline{\text{S}_1\text{P}}-\overline{\text{OP}}
    &= - \frac{l}{2}\cos\theta', \\
    \overline{\text{S}_2\text{P}}-\overline{\text{OP}}
    &= + \frac{l}{2}\cos\theta', \\
    \overline{\text{S}_3\text{P}}-\overline{\text{OP}}
    &= + \frac{l'}{2}\cos(\theta'-\pi/2). 
\end{align}
Thus the location of the amplitude zeros
depends
not only on $\pm\cos\theta'$
but also on the phase-shifted $+\cos(\theta'-\pi/2)$.
\footnote{
The existence of such a slit-like structure
within a scattering region
is a sign of chaos.
The chaoticity of string scatterings
will be discussed in the authors' next paper in preparation \cite{toappear}.
}

From these observations,
we can provide an interpretation that
the location of slits
are the end points of a string.
Firstly,
we should remark that
the order of the slit separation
is not the random walk size
$l\sim M^{1/2} \sim N^{1/4}$,
but rather
it is the total length of the string
$L\sim M \sim N^{1/2}$.
It motivates us
to assume that
the string
behaves like a simple harmonic oscillator 
in the scattering processes.

Then there remain two possibilities:
the locations of slits
indicate
(i) the nodes,
or (ii) the end points,
of the vibrating string.
The first possibility is excluded
since the HES amplitude with $J=1, N \gg 1$
behaves like a double-slit as discussed in Sec.~\ref{sec:HES}.
If the nodes of the vibrating string are the origin of 
the slit image,
the HES image with $J=1, N \gg 1$ would have been a multi-slit,
rather than the double-slit.
Since this is not the case, thus we are left with the possibility (ii): slits are end points of the string.

Such an interpretation
is also
supported by the following scenarios.
Suppose there is a classical vibrating string
with markers attached to its end points,
as an analogy of the world sheet with open string vertex operators inserted at its boundary.
The long time average
of the probability
that the markers are found at a point $X$
is roughly evaluated by
\begin{align}
    \int_{0}^{T} \frac{\dd{t}}{T}\:
    \delta(X-X_0\sin\omega t)
    \sim
    \frac{1}{\sqrt{X_0^2-X^2}}.
\end{align}
Then it becomes maximum
at the point
where the string is fully extended
to its maximum size.
This is consistent
with our results
where slits appeared
with the slit separations of the order of the total string length.

Another scenario
is about the origin
of interference patterns.
Suppose that incoming
two strings are shot along the $z$-direction
and scattered at the origin.
Our results have shown that
if this scattering
is observed from the $x$-direction,
the image is the double-slit
lined up along the $z$-direction.
It would be natural to
imagine that
a string formed at the origin
vibrates along the $z$-direction,
and is torn apart when the string reaches its maximum length.
Each of two pieces of the broken string can be scattered in any directions, 
though one must go in the opposite direction to the other. 
The observer in the $x$-direction can detect 
the scattered string with a definite probability. 
At the observation, the observer has two possibilities:
the left one of the two pieces of the broken string
reaches the observer 
while the right one
goes in the opposite direction, 
or visa versa.
The superposition
of these two possibilities
would be the source of the interference patterns which appear in the string scattering amplitudes.

These scenarios encourage us to state that the slits are the end points of a fundamental string.

\subsection*{Acknowledgment}

The work of K.~H.\ was supported in part by JSPS KAKENHI Grant No.~JP22H01217, JP22H05111 and JP22H05115.
The work of Y.~M.\ was supported in part by JSPS KAKENHI Grant No.~JP20K03930, JP21H05182 and JP21H05186. 
The work of T.~Y.\ was supported in part by JSPS KAKENHI Grant No.~JP22J15276.
The work of K.~H.\ and Y.~M.\ was also supported in part by JSPS KAKENHI Grant No.~JP17H06462.
K.~H.\ would like to thank Yannick Paget for discussions and collaborative artwork concerning this research.


\bibliographystyle{utphys}
\bibliography{ref}

\end{document}